**Inverse flexoelectret effect: bending dielectrics by a uniform electric field**

Xin Wen[1], Kai Tan[1], Qian Deng[1]*, Shengping Shen[1]*.

Key Laboratory for Strength and Vibration of Mechanical Structures, School of Aerospace Engineering, Xi'an Jiaotong University, Xi'an 710049, China.

***Corresponding author**: tonydqian@mail.xjtu.edu.cn (Q.D.); sshen@mail.xjtu.edu.cn (S.S.)

(Abstract) It is highly desirable to discover an electromechanical coupling that allows a dielectric material to generate curvature in response to a uniform electric field, which would add a new degree of freedom for designing actuators. Flexoelectricity, a two-way coupling between polarization and strain gradient, is a good candidate. But its applications are usually limited to the nanoscale due to its inherent size dependence. Here, an inverse flexoelectret effect in silicone elastomers is introduced to overcome this limitation. Based on this idea, a flexing actuator which can generate large curvature at the millimeter length scale is fabricated and shown to have excellent actuation performance comparable with current nanoscale flexoelectric actuators. Theoretical analysis indicates that the new phenomenon originates from the interplay of electrets and Maxwell stress. This work opens an avenue for applying macroscopic flexoelectricity in actuators and flexible electronics.

Usually, a uniform electric field can cause a single electro-active material which is homogeneous and isotropic deform uniformly via electromechanical couplings such as piezoelectricity [1], electrostriction [2], and Maxwell stress effect [3,4], which is widely used in actuations such as robotics [5,6] and artificial muscles [7]. To bend such homogeneous and isotropic active materials, a common way is to apply asymmetric constraints to it. For example, attaching a piece of active material onto the upper or lower surface of a passive cantilever beam which doesn't respond to the applied electric field can convert the uniform deformation into bending. This idea of generating curvature indirectly has been intensively applied in all kinds of actuators for decades [5-8] due to the advantages of bending deformation [9,10]. However, how to directly induce curvature by a uniform electric field, termed here as a "flexing actuator", remains a challenging problem.

A good candidate may be flexoelectricity, a universal and direct coupling between strain gradient (such as bending) and polarization in dielectrics [11-13]. Flexoelectricity is a two-way electromechanical coupling. In the past decade, direct flexoelectricity (polarization induced by strain gradient) has shown its great potential in applications including information storage and reading [14-16], tuning electronic properties [17-19], flexible electronics [20], bone repairing [21], energy harvesting [22-24], etc. On the other hand, converse (uniform strain induced by the gradient of electric field) and inverse (curvature induced by uniform electric field) flexoelectricity, although reported both theoretically [25-27] and experimentally [28-35], are still in their infancy. Like the direct flexoelectric effect, an inverse flexoelectric effect also shows strong size dependence [33]. Bursian et al. proposed that, because of the inverse flexoelectric effect, the induced curvature $\kappa$ of a dielectric cantilever beam was related to the uniformly applied electric field $E_V$ by [11]

$$\kappa = 12(1-\nu^2)\frac{\mu_{12}E_V}{Yh^2}, \qquad (1)$$

where $\mu_{12}$ is the transverse flexoelectric coefficient, $\nu$ denotes Poisson's ratio, $Y$ is Young's modulus, and $h$ is the thickness of the beam. With the decrease of $h$, the induced curvature $\kappa$ increases fast for a fixed applied electric field. Thus, as shown in Fig. 1(a), inverse flexoelectricity reduces sharply with the increase of the sample's feature size. The size effect seems to be an inevitable constraint to the application of both direct and inverse flexoelectricity at the macroscale.

Although several recent works show that the bending-induced polarization can be enhanced by 1~3 orders of magnitude through introducing extrinsic mechanisms [36-38], there is no evidence that these extrinsic mechanisms can naturally lead to an enhanced inverse effect. For the barrier-layer-mechanism, the most famous one among the above mechanisms, the electric-field-induced bending may be even weaker than its counterpart of intrinsic flexoelectricity [38]. By now, to our knowledge, there is no effective method to enhance inverse flexoelectricity to the macroscale, which is one of the most pressing challenges in the field. Very recently, we have reported another enhanced extrinsic mechanism (flexoelectret effect) to enhance the bending-induced polarization in silicone rubber with an embedded charge layer (flexoelectrets) [39]. One key feature to distinguish flexoelectret effect from other existing extrinsic mechanisms is that the flexoelectrets, with Young's modulus ~$10^2$ KPa, is about 6 orders of magnitude softer than typical flexoelectrics, which may suggest a potentially enhanced inverse phenomenon according to equation (1).

In this letter, we aim to explore (i) whether the flexoelectret effect is indeed a two-way electromechanical coupling, and (ii) whether the inverse effect (if it exists) is significant enough to trigger macroscopic applications. The central idea about the potential inverse flexoelectret effect is illustrated in Fig. 1(b) and (c): one layer of negative charges (red spheres with minus signs) is embedded in an elastomer beam, resulting in positive inductive charges (red spheres with plus signs) on surface electrodes and a pre-existing electric field (red arrows) which is symmetric about the charged plane. When an external voltage is applied, the external charges (yellow spheres) and inductive charges superpose on surface electrodes and results in larger electric field in the upper part while smaller electric field in the lower part (Fig. 1(c)). Hence, a macroscopic symmetry breaking of stretching in the sample, induced by Maxwell stress effect, is expected and leads to curvature eventually. This bending deformation would be further amplified due to the small Young's modulus of the material.

To test our idea, we fabricated a flexoelectret bar and clamped it on one end to form a flexing actuator (see the Supplemental Material [40]). Fig. 2(a) illustrates a brief schematic experimental setup for measuring the voltage-induced curvature: a laser displacement sensor was used to detect the displacement of the cantilever end generated by sinusoidal voltage from a high voltage supply (see the Supplemental Material [40]). Since the elastomer is non-polar, piezoelectricity is excluded here to explain any possible deformations induced by voltage (see the Supplemental Material [40]). The flexing actuator embedded with a charge layer in a density of -0.088mC/m$^2$ was stimulated by a sinusoidal voltage (1Hz) with a peak of 1kV. As shown in Fig. 2(b), sinusoidal bending oscillation (red circles) was observed, which shows the same frequency as the applied voltage (blue lines) and is nearly antiphase with the voltage signal. The output curvature, calculated from the P-P value of the Fourier-filtered first-harmonic displacement, varies linearly with the applied electric field (Fig. 2(c)). This linear coupling between curvature and electric field displays typical inverse flexoelectric characteristics, which is the first time that flexoelectric bending was observed at the macroscale.

To exam how much the intrinsic inverse flexoelectricity contributes to the bending of flexing actuators, a similar elastomer cantilever but without the charged layer was set to be a control group and tested at the same voltage stimulation. As shown by the black circles in Fig. 2(b), without the charge layer, only very tiny second-harmonic oscillations were observed, which we propose to be the uniform stretching of elastomer under voltage. The absence of first-harmonic responses can be explained by the fact that the intrinsic flexoelectricity is too weak to be observed at the macroscale as we mentioned above. Therefore, the bending deformations observed here originate from the charge layer in flexoelectrets rather than the distortion of material's microstructures (intrinsic mechanism) [28]. According to equation (1), the effective flexoelectric coefficient of flexoelectret (with the -0.088mC/m$^2$ charge layer) is $\mu_{12}^{eff} = 10.3\, nC/m$, comparable to the most famous flexoelectric polymer polyvinylidene difluoride (PVDF) [41]. Note that the flexoelectret introduced here is much softer than PVDF (Young's modulus is more than 4 orders of magnitude smaller), which results in much larger bending deformation than that of PVDF according to equation (1).

To further confirm the effect of the charge layer on this pronounced output curvature, we first replaced the negative charges of the middle layer with positive charges, which was performed by changing the voltage polarity of the needle tip in corona charging process. The flexing actuator with a positive charge layer (0.086mC/m$^2$) was stimulated by the same sinusoidal voltage (1Hz in frequency and 1kV in amplitude), whereas the bending oscillation signal here is nearly in phase with the driving voltage (Fig. 2(d)), and exactly opposite to the case in Fig. 2(b). This means that the bending direction of a flexing actuator depends on the polarity of the embedded charge layer. Then, we fixed the polarity (negative) of the charge layer and further changed the magnitude of the charge density. As shown in Fig. 2(e), the output curvature increases almost linearly with the magnitude of charge density. These observations not only provide strong evidence that the enhanced inverse flexoelectret effect is related to the embedded charge layer, but also shows that the effect is tunable via manipulating the charge layer. Nevertheless, how the charge layer influences the electromechanical coupling still needs further exploration.

To identify the main cause of the enhanced inverse flexoelectric phenomenon, we establish an electrostatic model (Fig. 3(a)) in which the only electromechanical coupling mechanism is the Maxwell stress effect. A charge layer with a density of $q$ is set on the middle plane. Note that although the flexoelectrets were fabricated by placing a charged polytetrafluoroethylene (PTFE) thin film into elastomers in the experiments (see the Supplemental Material[40]), the possible effect of the PTFE layer on the electromechanical response of flexoelectrets is minimized due to its negligible thickness ($10\mu m$) compared to the elastomer matrix (see the Supplemental Material [40] for laminated model and detailed discussion). To illuminate the physical mechanism more clearly, we treat the charged PTFE film as a charge layer with no thickness in this theoretical model. The electric field throughout the model comes from the charge layer (red spheres) and external voltage, which can be calculated from Gauss' law (see the Supplemental Material [40]) as

$$E^a = \frac{q}{2\varepsilon} - \frac{V}{h}, E^b = -\frac{q}{2\varepsilon} - \frac{V}{h}, \qquad (2)$$

where $q$ is the density of the charge layer, $\varepsilon$ is the permittivity, $E^a$ and $E^b$ denote the electric field above and below the charge layer respectively. Assuming that the thickness $h$, charge density $q$ and permittivity $\varepsilon_r$ are $0.8mm$, $-0.1mC/m^2$ and $2.85\varepsilon_0$ respectively, the electric field (absolute value) along thickness direction as a function of the applied voltage can be plotted in the rainbow diagram of Fig. 3(b). When no voltage is applied, $E^a$ and $E^b$ become $q/2\varepsilon$ and $-q/2\varepsilon$ respectively, showing the same magnitude but opposite directions as illustrated in Fig. 3(a) and (b). At the excitation of voltage, the electric field is no longer symmetric about the middle plane. The lines with circles and triangles (Fig. 3(b)) depict the change of electric field above and below middle plane as a function of voltage. If the voltage is positive, then the upper part would experience an electric field (absolute value) larger than the lower part as shown in Fig. 3(c).

Further, dielectric material would experience Maxwell stress due to the electric field it feels. Among the three components of Maxwell stress ($p_1$, $p_2$, $p_3$), it's the stretching stress along cantilever's axis ($p_2$) that determines the cantilever bending, which can be expressed as

$$p_2^a = \frac{1}{2}\varepsilon(E^a)^2 = \frac{1}{2}\varepsilon(\frac{q}{2\varepsilon} - \frac{V}{h})^2 \qquad (3)$$

and

$$p_2^b = \frac{1}{2}\varepsilon(E^b)^2 = \frac{1}{2}\varepsilon(\frac{q}{2\varepsilon} + \frac{V}{h})^2, \qquad (4)$$

where $p_2^a$ and $p_2^b$ denote the stress $p_2$ above and below the charge layer respectively. Similarly, the response of Maxwell stress to applied voltage (Fig. 3(d)) mirrors the situation in Fig. 3(b) with a difference that stress varies nonlinearly with voltage, which is the typical nonlinear characteristic of Maxwell stress[4,7]. Given that one part of the model undergoes larger stretching stress than the other part, a resultant force $F$ plus a bending moment $M$ can be obtained as

$$F = b \int_{-h/2}^{h/2} p_2 dx_1 = \frac{bhq^2}{8\varepsilon} + \frac{\varepsilon bh E_V^2}{2} \qquad (5)$$

and

$$M = b \int_{-h/2}^{h/2} x_1 p_2 dx_1 = -\frac{qbE_V h^2}{8}, \qquad (6)$$

where $b$ is the width of the model. Interestingly, in the above two equations, the square term of the applied electric field $E_V^2$ remains in the former one but is eliminated in the latter one, which implies a linear coupling between curvature $\kappa$ and external electric field $E_V$ that can be obtained as

$$\kappa = \frac{M}{YI} = -\frac{3}{2}\frac{qE_V}{Yh}, \qquad (7)$$

where $I = bh^3/12$ is the bending stiffness which links the bending moment to the induced curvature of a beam. This formula predicts that the bending direction depends on the polarity of the charge density and applied voltage, consistent with our experimental observations. We also performed finite element calculations considering Maxwell stress effect to solve this problem (see the Supplemental Material[40]). We substitute the experimentally measured material properties into equation (7) and FEM model. As shown in Fig. 3(e) and (f), our theoretical and simulation results show good agreements with experimental results for samples with different charge densities or thickness. Thus, the origin of the enhanced curvature can be understood: due to the interplay with electrets, the Maxwell stress that varies quadratically to the applied voltage is coaxed to an inverse flexoelectret effect that linearly couples the applied voltage and bending deformation.

The actuation performance evaluated by curvature/electric field ratio for flexoelectrets and typical flexoelectrics like SrTiO$_3$[15c, 15e] and BaTiO$_3$ [15a] are compared In Fig. 4(a). It can be seen that flexoelectrets become the first flexoelectric member that enables a macroscopic flexing actuator, which is not only due to the soft materials it composed of, but, more importantly, also because of the nature of the inverse flexoelectret effect itself. For intrinsic inverse flexoelectricity, as shown by equation (1), the induced curvature is proportional to $1/h^2$.

However, for the inverse flexoelectret effect introduced here, as shown by equation (7), the induced curvature is proportional to $1/h$. This difference results in a quite different size dependency of them, which is schematically plotted in Fig. 4(b). Under a fixed external electric field, the bending angle of a traditional flexoelectric actuator decreases fast with the increase of the sample size. While for the beam showing flexoelectret behaviors, the bending angle of the beam is independent of the sample size. Thus, the inverse flexoelectret effect we introduced here breaks the constraint of size effect that limits the applications of flexoelectricity to the nanoscale, just as the schematic illustration in Fig. 1.

In summary, this work demonstrates that size scaling does not necessarily result in deterioration of inverse flexoelectricity at the macroscale. Different from the strong size dependence of intrinsic flexoelectricity, applying a fixed electric field to dielectrics can induce a size-independent bending behavior through the inverse flexoelectret effect, which is constructed by the interplay of electrets and Maxwell stress. This enables a macroscopic flexing actuator capable of generating curvature in response to a uniform electric field, of which the actuation performance is comparable with the nanoscale flexoelectric actuator. Looking beyond silicone rubber and elastomer, inverse flexoelectret effect should in principle exist in all dielectrics because of the universality of Maxwell stress effect, suggesting a wide range of materials for fabricating flexing actuators. In the meantime, this work also provides a solution for electrets to work effectively in actuating mode, which is usually not allowed for the famous piezoelectrets due to its relatively hard polymer matrix (0.1~1GPa) and/or poor performance in bending deformations [42-44]. In a word, our work adds a new degree of freedom for designing actuators[45] and flexible electronics[46], which is helpful to simplify device design, avoid interface failure and may achieve novel functional properties that are previously inaccessible. For example, flexoelectrets may serve as a platform in which direct/inverse flexoelectret effect can be coupled to other functional properties such as ferroelectricity[14], magnetism[47] and semiconductor characteristic[17], since these properties can be easily introduced to silicone elastomer/rubber[48-50].

**Acknowledgments:**

We gratefully acknowledge the support from the National Key R&D Program of China (2017YFE0119800), the National Natural Science Foundation of China (Grant Nos. 11632014, 11672222, and 11372238), the 111 Project (B18040), the Chang Jiang Scholar Program, and numerous helpful discussions with Jingran Liu (Xi'an Jiaotong University).

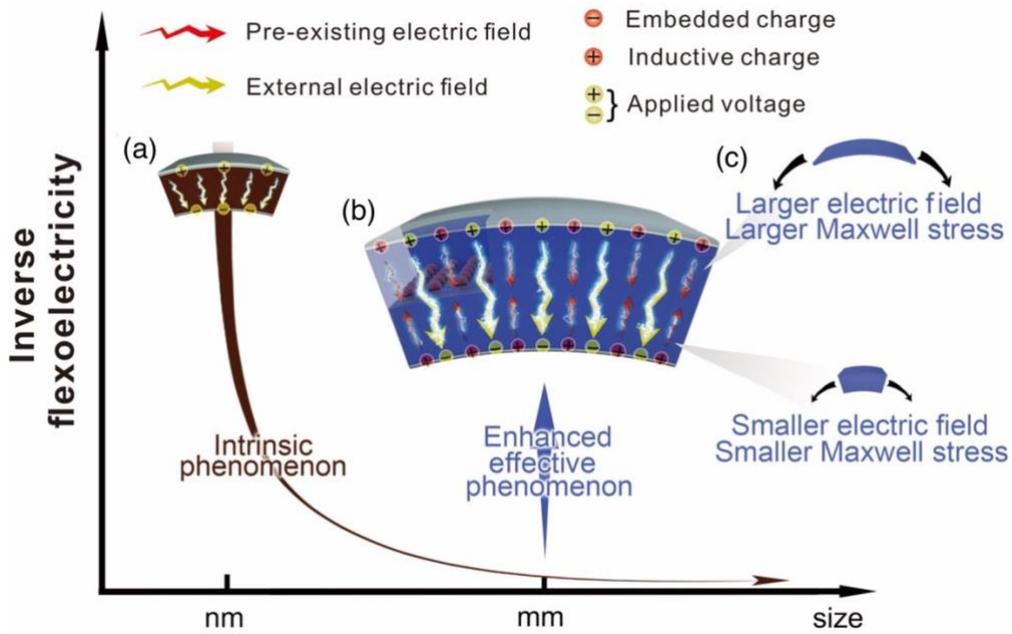

**FIG 1. Schematic illustration of the mechanism of inverse flexoelectret effect.** (**a**) An external electric field (yellow arrows) generated by external voltage (yellow spheres) is capable of bending a dielectric material via intrinsic flexoelectricity, which is significant at the nanoscale while decreases dramatically with size increases. (**b**) The effective flexoelectricity is enhanced in elastomer with a layer of embedded negative charges (red spheres with minus signs). The embedded charges induce positive charges on surface electrodes (red spheres with plus signs) and consequent pre-existing electric field (red arrows). Under external voltage, the external charges (yellow spheres) and inductive charges add to each other on the upper electrode while canceling on the other one. (**c**) Thus, upper and lower parts undergo different true electric field and consequent different stretching stress via Maxwell stress effect, leading to bending deformation.

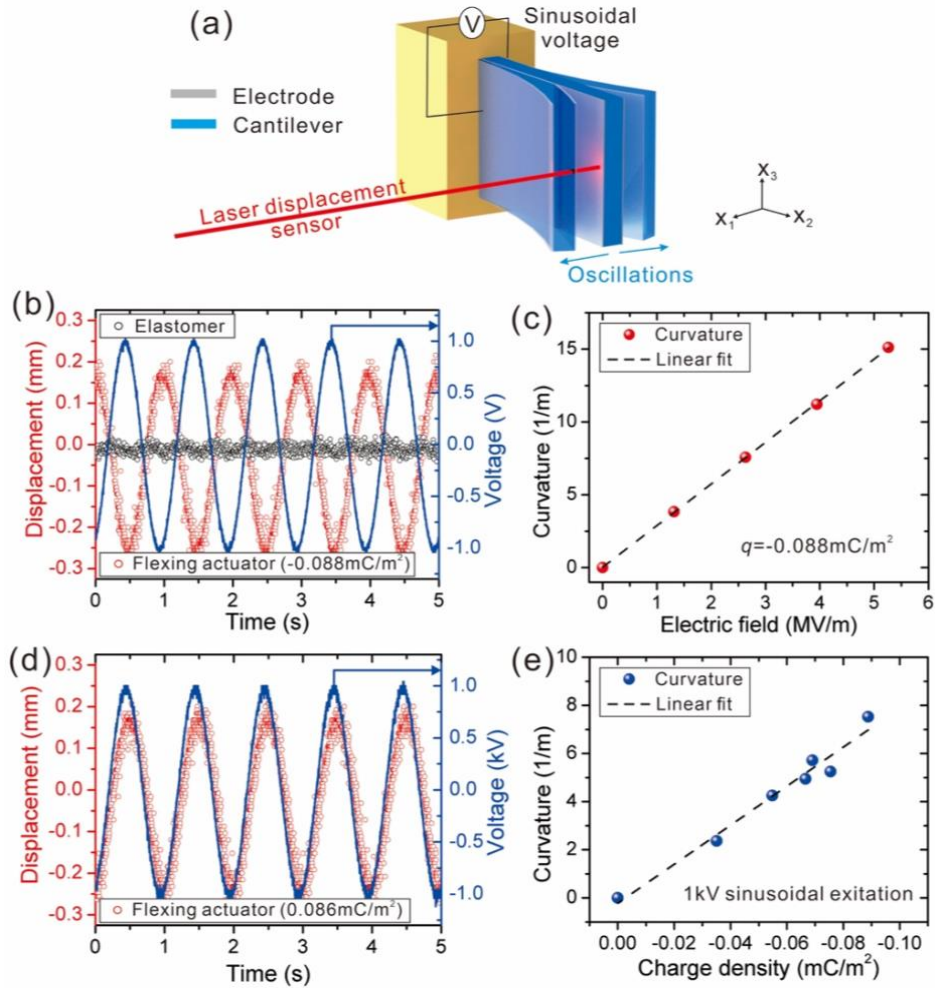

**FIG 2. Experimental setup and results. (a)** Experimental setup consisting of a clamping rigid body, a high voltage supply, and a laser displacement sensor. **(b)** Applied voltage (blue solid line) and induced displacement of cantilever end. Red and black circles represent elastomer without charge and flexing actuator with a charge density of -0.088mC/m², respectively. **(c)** Curvature as a function of the applied electric field for flexing actuator (-0.088mC/m²). **(d)** Applied voltage (blue solid line) and induced first-harmonic displacement of flexing actuator cantilever end. The charge density of the middle layer is positive (0.086mC/m²). **(e)** Curvature as a function of the magnitude of charge density under the bias of 1kV. All results in this figure were acquired for the sample with a size of 10mm7mm0.76mm at sinusoidal excitation of 1Hz at room temperature.

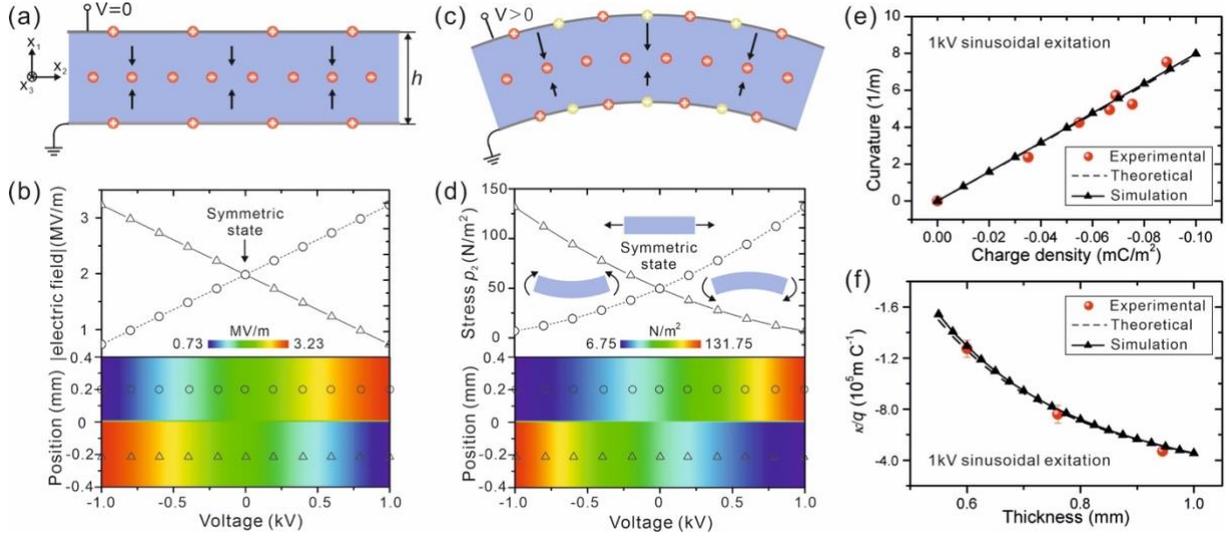

**FIG 3. Electrostatic model of a flexoelectret and comparison between experimental, theoretical and simulation results.** (a) Theoretical model of a flexoelectret. The red spheres with minus signs represent the embedded negative charges with a density of $q$. The red spheres with plus signs represent the inductive positive charges on electrodes. The black arrows represent the electric field. (b) Distribution of the electric field (absolute value) along thickness direction as a function of the applied voltage. The lines with circles and triangles represent the electric field above and below middle plane respectively. (c) Schematic illustration of electric field inside the model when applied voltage is positive. The yellow spheres represent applied voltage. (d) Distribution of Maxwell stress $p_2$ along thickness direction as a function of the applied voltage. **(e)** and **(f)** Curvature as a function of charge density (e) and thickness (f) and the comparison of experimental, theoretical, and simulation results. To only investigate the influence from thickness change, we divide the curvature $\kappa$ by charge $q$ because it is difficult to control charge density accurately.

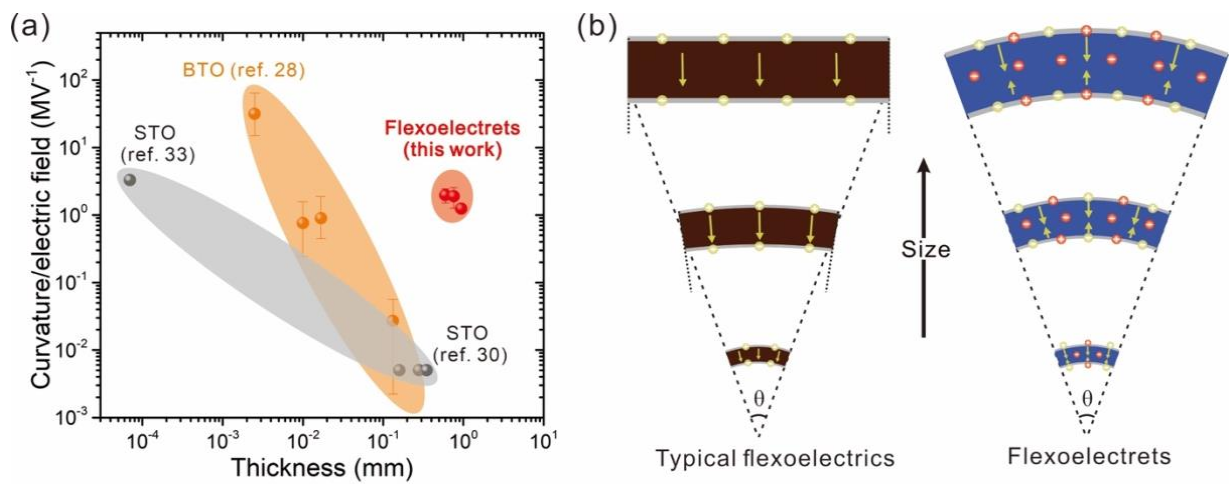

**FIG 4. Comparison of the bending performance between flexoelectrets and typical flexoelectrics.** **(a)** The ratios of curvature/electric field as a function of thickness are compared for flexoelectrets (this work) and flexoelectric $SrTiO_3$ [30,33] and $BaTiO_3$ [28]. **(b)** The evolution of bending performance under a fixed applied electric field (yellow arrows) when the size is scaled up.

# Supplemental Material for
# ¨Inverse flexoelectret effect: bending dielectrics by a uniform electric field¨

X. Wen[1], K. Tan[1], Q. Deng[1]*, S. Shen[1]*.

**Affiliations:** [1]State Key Laboratory for Strength and Vibration of Mechanical Structures, School of Aerospace Engineering, Xi'an Jiaotong University, Xi'an 710049, China.

***Corresponding author**: tonydqian@mail.xjtu.edu.cn (Q.D.); sshen@mail.xjtu.edu.cn (S.S.)

Material fabrication

Fig. S1 illustrates the fabrication process of elastomer without charge layer (control group) and flexoelectrets with a charge layer (experimental group). The parts A and B of ecoflex 0010 were thoroughly mixed in a ratio of 1:1. Then, the mixture was transferred onto an acrylic substrate (50×50mm$^2$) using the spin-coating method (KW-4A, SETCAS). The elastomer with different thicknesses (0.3mm/0.38mm/0.472mm) was obtained by controlling the rotate speed and time. In the next step, the acrylic substrate with a layer of the mixture was put in an oven at 60℃. After 4 hours, the solidified elastomer was peeled off and cut to a specific size (11mm×7mm×0.3mm/0.38mm/0.472mm). Subsequently, we put a polytetrafluoroethylene (PTFE) thin film with a thickness of 10 $\mu$m onto one surface of an elastomer plate, followed by another elastomer plate with the same thickness. Then, the control group without charge layer was obtained as shown in Fig. S1. Note that the adhesion between film and elastomer was completed by the strong stickiness of the elastomer itself. Also, to reduce the possible limitation of PTFE film on the deformation of elastomer, the film was cut into strips before we put it into the elastomer.

To obtain flexoelectrets, we performed corona charging treatment to PTFE film before the top elastomer was put on the sample as shown in Fig. S1. The film was placed 10cm below the corona needle tip that was kept to the voltage of -15 kV for 3mins using a high voltage power supply (DW-SA303-3ACF1, Dongwen). Finally, we put another elastomer onto the charged surface and obtained flexoelectrets with a charge layer in the middle plane (experimental group). Compliant electrodes (conductive silicone grease) were applied to the upper and lower surfaces of all samples for measurements of voltage-induced oscillation.

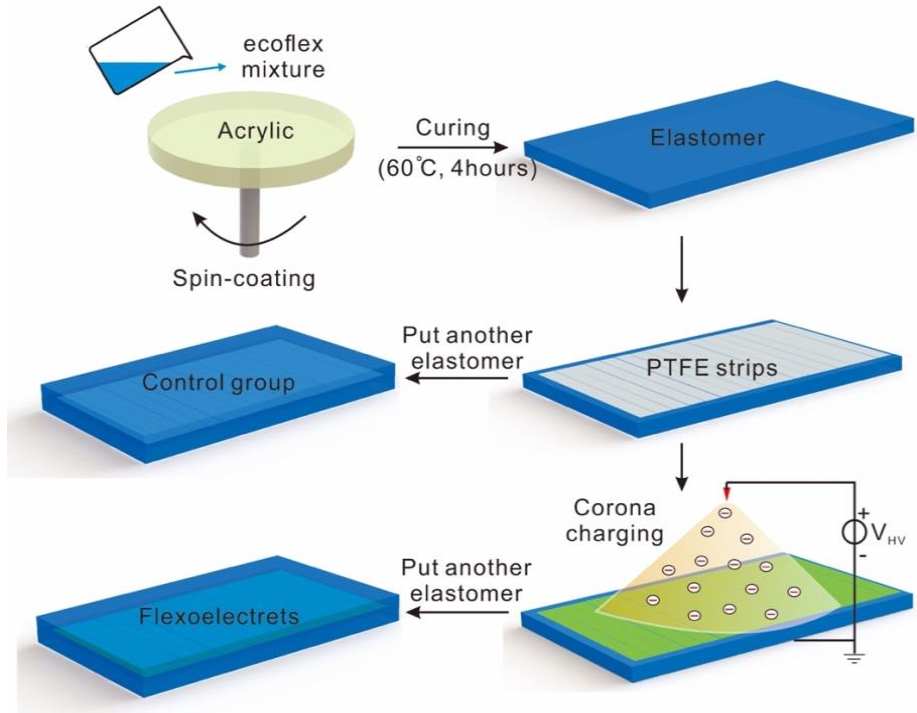

**FIG. S1.** Fabrication of elastomer (control group) and flexoelectrets (experimental group).

Material characterization

A commercial piezometer (PM-200, Piezotest) with a revolution of 0.01pC/N was used to measure the piezoelectric coefficient $d_{33}$ and no effective signal was observed, showing that $d_{33}$ ＜0.01pC/N and thus our elastomer is not piezoelectric. Young's modulus was measured by a load machine (electroforce 3230, TA) using an elastomer with a size of 19mm×6mm×0.472mm. The stress-strain curve was shown in Fig. S2 and Young's modulus is equal to the slope value ($Y$=56.79 KPa). An elastomer block with a cube shape (10mm×10mm×10mm) was prepared to measure the density by dividing the mass by the volume (ρ=1148.5kg/m³). Charge density $q$ was measured by an electrostatic voltmeter (ModelP0865, Trek). After the corona charging process shown in Fig. S1, we used the voltmeter to measure the surface potential $U$ of the charged film, which could be converted to charge density by $q = \varepsilon U/h$ where $\varepsilon$ and $h$ is permittivity and half the thickness of our final sample respectively [1]. An impedance analyzer (E4990A, Keysight) was used to measure the relative permittivity $\varepsilon_r = 2.85$. All these measurements were performed at room temperature.

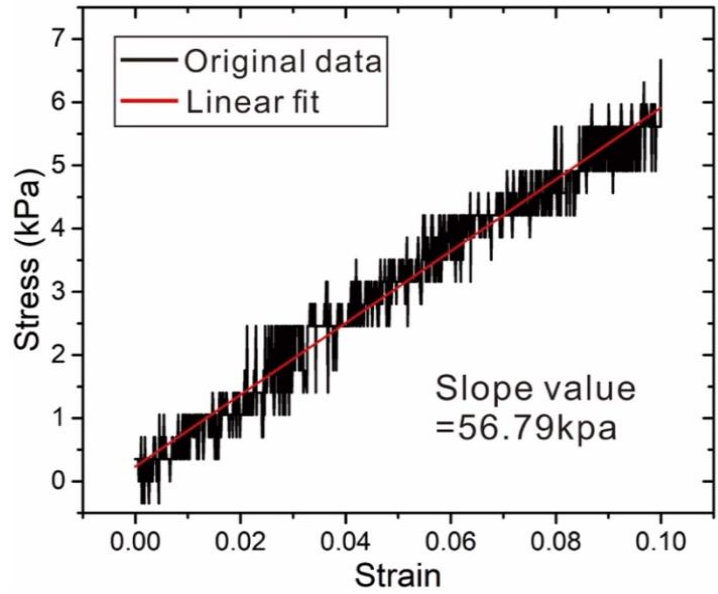

**Fig. S2.** Stress-strain curve of the elastomer for measuring Young's modulus.

Experimental setup

Fig. S3 illustrates the experimental setup for measuring voltage-induced oscillations. To reduce the noise from the environmental vibration, the measurement was performed at an environmental vibration isolating desk (Vision IsoStation, Newport). The sample with electrodes on two surfaces was clamped by two blocks of Polyvinylidene Fluoride. An AC power supply (AMS-10B2-L, Matsusada) was used to apply sinusoidal voltage on the cantilever. The ground wire (black clip connector) and the voltage wire (red clip connector) was connected to the left and the right electrodes through conductive tapes and copper wires. A laser displacement sensor (LK-H025, Keyence) was used to detect the displacement of the cantilever end. In fact, as shown in the inset of Fig. S3, the laser spot is about 0.5mm from the end to avoid the laser scattering. The voltage signal and the displacement signal were recorded simultaneously by an oscilloscope (MDO3104, Tektronix).

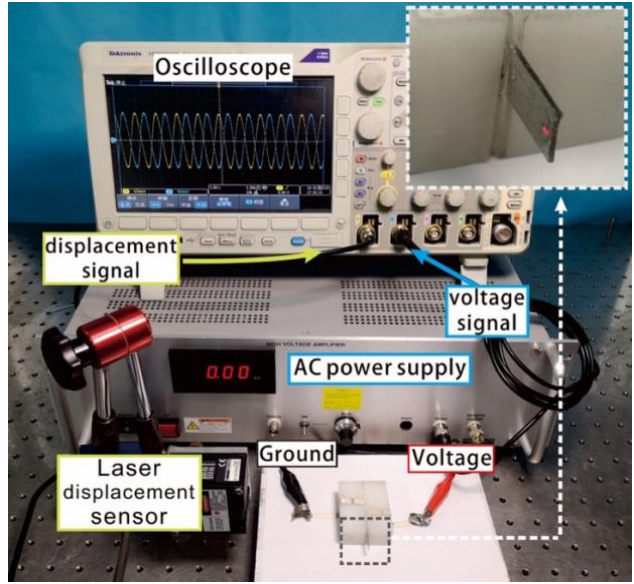

**FIG. S3.** Experimental setup for measuring the voltage-induced oscillations, consisting of an environmental vibration isolating desk, a laser displacement sensor, an AC power supply, and an oscilloscope. The inset shows the enlarged view of the cantilever.

Calculation of the electric field distribution

In the model shown in Fig. 3(a), the electric potential of the bottom electrode is set to be 0 and we apply voltage $V$ to the top electrode. According to Gauss' law, it is easy to obtain that the electric field distribution in the model is uniform above or below the charge layer, which is marked as $E^a$ and $E^b$ respectively. The charge density of upper and lower electrodes is $q_1$ and $q_2$ respectively. Applying Gauss' law to the model, we have

$$\begin{cases} 0 - E^a = q_1/\varepsilon \\ E^a - E^b = q/\varepsilon \\ E^b - 0 = q_2/\varepsilon \end{cases} \tag{S1}$$

The electrical boundary condition can be written as

$$E^a \frac{h}{2} + E^b \frac{h}{2} = V \tag{S2}$$

Using equation (S1) and (S2), we obtain the electric field distribution shown in equation (2) in main text.

Derivation of the voltage-curvature relationship in a laminated model

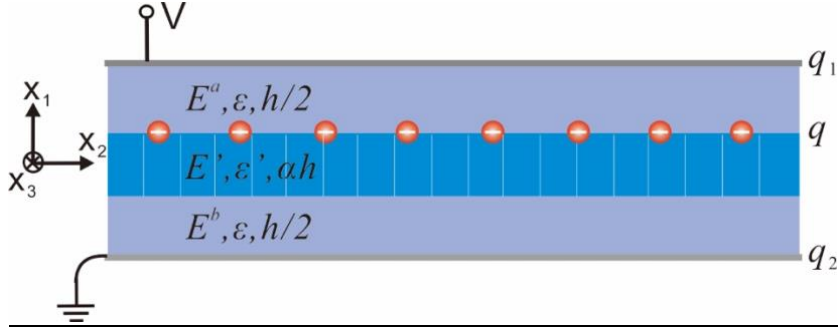

**FIG. S4.** Theoretical laminated model of a flexoelectret considering the PTFE layer. The red spheres with minus signs represent the embedded negative charges with a density of $q$.

To determine how would the possible asymmetry introduced by the fabrication method affect inverse flexoelectret effect, we consider a laminated model consisting with one layer of PTFE and two layers of elastomers (Fig. S4). The charge layer is placed on the upper surface of PTFE part. The white vertical line means that the PTFE layer is aligned in the form of strips as shown in Fig. S1, so the PTFE layer does not affect the stretching deformation of the elastomer layers. The thickness, Young's modulus, and permittivity of two elastomer layers and PTFE layer are $h/2, Y, \varepsilon$ and $\alpha h, Y', \varepsilon'$ respectively. The electric field distributed in three layers are marked as $E^a, E'$, and $E^b$ respectively. Applying Gauss' law to the model, we have

$$\begin{cases} 0 - \varepsilon E^a = q_1 \\ \varepsilon E^a - \varepsilon' E' = q \\ \varepsilon' E' - \varepsilon E^b = 0 \\ \varepsilon E^b - 0 = q_2 \end{cases} \quad (S3)$$

The electrical boundary condition can be written as

$$E^a \frac{h}{2} + E'(\alpha h) + E^b \frac{h}{2} = V \quad (S4)$$

Using equation (S3) and (S4), the electric field distribution can be obtained as

$$E^a = \frac{\frac{V\varepsilon'}{h} + q(\alpha + \frac{\varepsilon'}{2\varepsilon})}{\varepsilon' + \alpha\varepsilon}, \; E' = \frac{\frac{V\varepsilon}{h} - \frac{q}{2}}{\varepsilon' + \alpha\varepsilon}, \; E^b = \frac{\frac{V\varepsilon'}{h} - \frac{q\varepsilon'}{2\varepsilon}}{\varepsilon' + \alpha\varepsilon} \quad (S5)$$

Since the Young's modulus of PTFE is much larger than the Young's modulus of the elastomer (almost 4 orders of magnitude), the deformation of the whole model is dominated by the two active elastomer layers. The Maxwell stress along $x_2$ direction in two elastomer layers can be written as

$$p_2^a = \frac{1}{2}\varepsilon\left[\frac{\frac{V\varepsilon'}{h} + q(\alpha + \frac{\varepsilon'}{2\varepsilon})}{\varepsilon' + \alpha\varepsilon}\right]^2, \quad p_2^b = \frac{1}{2}\varepsilon\left[\frac{\frac{V\varepsilon'}{h} - \frac{q\varepsilon'}{2\varepsilon}}{\varepsilon' + \alpha\varepsilon}\right]^2 \quad (S6)$$

The stress is equivalent to a resultant force $F$ and a bending moment $M$, which can be expressed as

$$F = \frac{bh\varepsilon}{4(\varepsilon' + \alpha\varepsilon)^2}\left[q^2\alpha^2 + \frac{2\alpha qV\varepsilon'}{h} + \frac{2V^2\varepsilon'^2}{h^2} + \frac{q^2\alpha\varepsilon'}{\varepsilon} + \frac{q^2\varepsilon'^2}{2\varepsilon^2}\right]$$

$$M = -\frac{(1+2\alpha)\varepsilon bh^2}{16(\varepsilon' + \alpha\varepsilon)^2}\left[q^2\alpha^2 + \frac{q^2\alpha\varepsilon'}{\varepsilon} + \frac{2\alpha qV\varepsilon'}{h} + \frac{2qV\varepsilon'^2}{\varepsilon h}\right] \quad (S7)$$

The bending stiffness of the laminated beam is

$$D = Y'\int_{-\alpha h/2}^{\alpha h/2} bx_1^2 dx_1 + Y\int_{-(1+\alpha)h/2}^{-\alpha h/2} bx_1^2 dx_1 + Y\int_{\alpha h/2}^{(1+\alpha)h/2} bx_1^2 dx_1$$

$$= \frac{bh^3\left[\alpha^3(Y' - Y) + (1 + \alpha)^3 Y\right]}{12} \quad (S8)$$

Dividing the bending monment by the bending stiffness, we obtain the curvature

$$\kappa = \frac{M}{D} = -\frac{3\varepsilon(1+2\alpha)\left[q^2\alpha^2 + \frac{q^2\alpha\varepsilon'}{\varepsilon} + \frac{2\alpha qV\varepsilon'}{h} + \frac{2qV\varepsilon'^2}{\varepsilon h}\right]}{4h(\varepsilon' + \alpha\varepsilon)^2\left[\alpha^3(Y' - Y) + (1 + \alpha)^3 Y\right]} \quad (S8)$$

If $\alpha \to 0$, the output curvature can be further reduced to a simplified equation (7) in the updated manuscript.

Equation (S8) determines the effect of PTFE layer on bending deformation. As shown in Fig. S5(a), the output curvature reduces as $\alpha$ becomes larger, and this reduction is asymmetric for applied voltage with opposite polarity. We further define the bending asymmetry coefficient as

$$\%\text{Asy} = \frac{|\kappa^+| - |\kappa^-|}{\langle\kappa\rangle} 100 \quad (9)$$

where $\kappa^+$ and $\kappa^-$ are the curvature under positive voltage and negative voltage respectively. The average curvature is $\langle\kappa\rangle = (|\kappa^+| - |\kappa^-|)/2$. As shown in Fig. S5(b), the asymmetry increases almost linearly with the relative thickness of the PTFE layer. Since the PTFE film is very thin (20$\mu$m), $\alpha$ in our experiments is between 0.011 to 0.017, corresponding to the hardly reduced curvature and tiny bending asymmetry less than 5%.

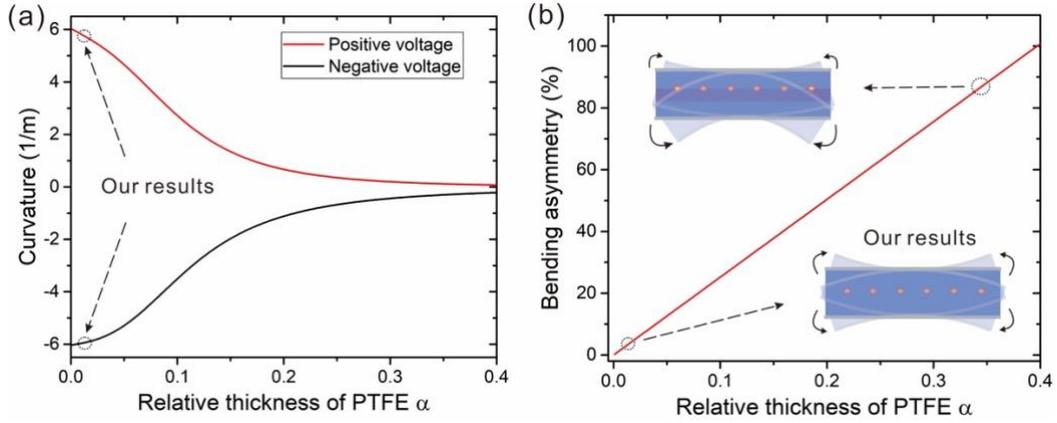

**FIG S5**. The effect of the material asymmetry on the inverse flexoelectret effect. (a) Output curvature and (b) bending asymmetry as a function of the relative thickness of PTFE. In our experiments, $\alpha$ is from 0.011 to 0.017, which is nearly the case without considering PTFE layer. The materials properties used in this figure are: $h = 0.76\text{mm}$, $q = -0.88\text{mC/m}^2$, $V = \pm 1.5\text{kV}$, $\varepsilon = 2.85$, $\varepsilon' = 2$, $Y = 56.79\text{KPa}$, $Y' = 0.4\text{GPa}$.

FEM calculations

Fig. S6(a) illustrates our FEM model for calculating the flexoelectrets' electromechanical response using COMSOL MULTIPHYSICS 5.2. We give the model material properties measured before: relative permittivity $\varepsilon_r = 2.85$, Young's Modulus $Y=56.79\text{KPa}$, density $\rho=1148.5\text{kg/m}^3$. Since silicone elastomers are nearly incompressible, Poisson's ratio $\nu$ is set to be 0.49. The material properties are uniform in the whole model and the model is isotropic. Same as the settings in the experiment as shown in Fig. 2(a) and Fig. S3, we set fixed constraints for one cantilever end and set the other end free. At the same time, we set the electric potential of one side to be 0V and change the electrical potential of the other side. Besides, different values of charge density are given to the middle plane as the red rectangular box shown in fig. S6(a). Note that, in our model, piezoelectric coefficient and electrostrictive coefficient are set to be zero. Also, COMSOL (actually, all current commercial software) does not include intrinsic flexoelectricity. Therefore, the only electromechanical coupling phenomenon considered in our model is the Maxwell stress effect, which means any deformation induced by voltage purely from Maxwell stress effect. Linear elastic model is given to the model and geometric nonlinearity is considered in the FEM calculating.

To show the calculated results clearly, the results of the $x_1$-$x_2$ cross-section are plotted in Fig. S6(b) to (d), which show the deformations of the model with a charge density of - 0.1$mC/m^2$, +0.1$mC/m^2$, and 0$mC/m^2$ when subjected to $\pm 1kV$ and the corresponding Maxwell stress $p_2$ distribution. In the initial state, the charge layer generated an electric field with the same magnitude across the model, leading to uniform Maxwell stress as shown in Fig. S6(b) and (c). At the excitation of the voltage, the symmetric state of stress was broken. Specifically, one part experienced larger stretching stress than the other part, which leads to a bending moment *M*. The direction of bending moment depends on the polarity of charge layer. For the model without charge layer as shown in Fig. S6(d), no bending deformation was generated at the same voltage stimulation. The curvature is calculated by dividing the difference of the average strain of two sides by the thickness of the model. The results in Fig. 3(d) (Fig. 3(e)) was obtained by changing the charge density (thickness) while keeping material properties and other settings unchanged. Specifically, the charge density is set to be from -0.01~-0.1mC/m² at the interval of -0.01mC/m², and thickness is set to be from 0.55mm~1mm at the interval of 0.025mm.

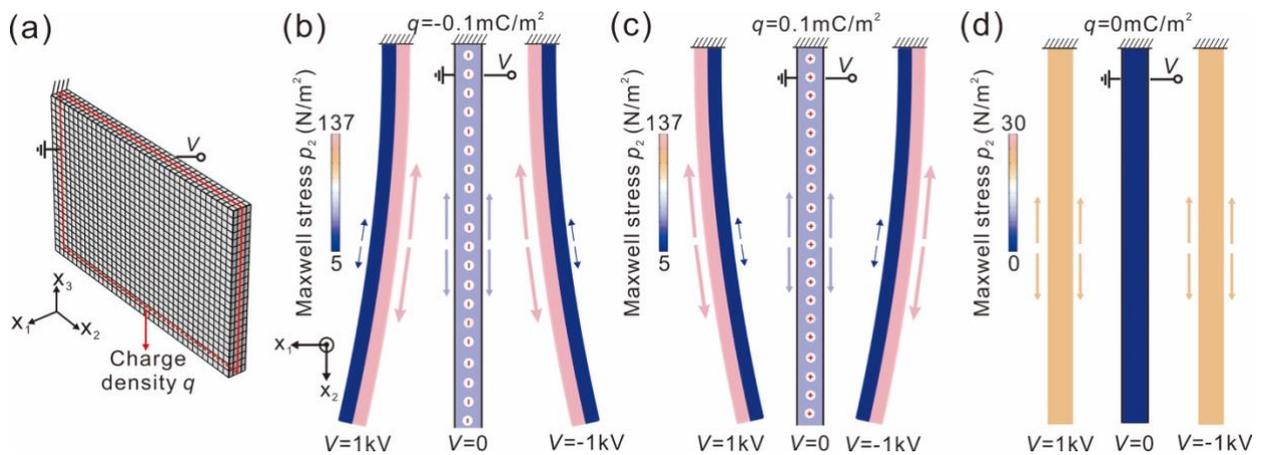

**FIG. S6.** The FEM model and calculating results. (**a**) The FEM model with a charge layer in the middle plane. (**b** to **d**) The distribution of Maxwell stress *p₂* for the model with a negative charge layer (b), a positive layer (c), and without chare layer (d) at the excitation of $\pm 1kV$. Arrows roughly depicted the magnitude and direction of *p₂*.

[1] R. Kacprzyk, E. Motyl, J. B. Gajewski, A. Pasternak, Piezoelectric properties of nonuniform electrets. *Journal of Electrostatics* **35**, 161-166 (1995).

23